\documentclass{elsart}
\journal{Physics Letter B}
\usepackage{graphics}
\usepackage{color}

\makeatletter
\DeclareRobustCommand*\cal{\@fontswitch\relax\mathcal}
\DeclareRobustCommand*\mit{\@fontswitch\relax\mathnormal}

\def\endequation{\eqno \hbox{\@eqnnum}$$\@ignoretrue}
\def\eqnarray{%
   \stepcounter{equation}%
   \def\@currentlabel{\p@equation\theequation}%
   \global\@eqnswtrue
   \m@th
   \global\@eqcnt\z@
   \tabskip\@centering
   \let\\\@eqncr
   $$\everycr{}\halign to\displaywidth\bgroup
       \hskip\@centering$\displaystyle\tabskip\z@skip{##}$\@eqnsel
      &\global\@eqcnt\@ne\hskip \tw@\arraycolsep \hfil${##}$\hfil
      &\global\@eqcnt\tw@ \hskip \tw@\arraycolsep
         $\displaystyle{##}$\hfil\tabskip\@centering
      &\global\@eqcnt\thr@@ \hb@xt@\z@\bgroup\hss##\egroup
         \tabskip\z@skip
      \cr
}
\def\endeqnarray{%
      \@@eqncr
      \egroup
      \global\advance\c@equation\m@ne
   $$\@ignoretrue
}
\makeatother

\begin{document}

\begin{frontmatter}

\begin{flushleft}
\hspace*{9cm}BELLE Preprint 2006-24 \\
\hspace*{9cm}KEK Preprint 2006-32 \\
\hspace*{9cm}hep-ex/0609018
\end{flushleft}

\title{
First Observation of the Decay $\tau^- \to \phi K^- \nu_\tau$
}

\collab{Belle Collaboration}
  \author[Nagoya]{K.~Inami}, 
  \author[KEK]{K.~Abe}, 
  \author[TohokuGakuin]{K.~Abe}, 
  \author[KEK]{I.~Adachi}, 
  \author[Tokyo]{H.~Aihara}, 
  \author[BINP]{D.~Anipko}, 
  \author[BINP]{K.~Arinstein}, 
  \author[BINP]{V.~Aulchenko}, 
  \author[Lausanne,ITEP]{T.~Aushev}, 
  \author[Cincinnati]{S.~Bahinipati}, 
% ...

\address[BINP]{Budker Institute of Nuclear Physics, Novosibirsk, Russia}
% ...
\address[Cincinnati]{University of Cincinnati, Cincinnati, OH, USA}
% ...
\address[KEK]{High Energy Accelerator Research Organization (KEK), Tsukuba, Japan}
% ...
\address[ITEP]{Institute for Theoretical and Experimental Physics, Moscow, Russia}
% ...
\address[Lausanne]{Swiss Federal Institute of Technology of Lausanne, EPFL, Lausanne, Switzerland}
% ...
\address[Nagoya]{Nagoya University, Nagoya, Japan}
% ...
\address[TohokuGakuin]{Tohoku Gakuin University, Tagajo, Japan}
% ...
\address[Tokyo]{Department of Physics, University of Tokyo, Tokyo, Japan}
% ...

\begin{abstract}
We present the first observation of $\tau$ lepton decays to hadronic final 
states with a $\phi$-meson.
This analysis is based on 401 fb$^{-1}$ of data accumulated
at the Belle experiment. The branching fraction obtained is
${\cal B}(\tau^-\to\phi K^-\nu_\tau) = (4.05\pm 0.25\pm 0.26)\times 10^{-5}$.
\end{abstract}

\end{frontmatter}


\section{Introduction}

\begin{figure}[htb]
\centerline{\resizebox{3.5cm}{!}{\includegraphics{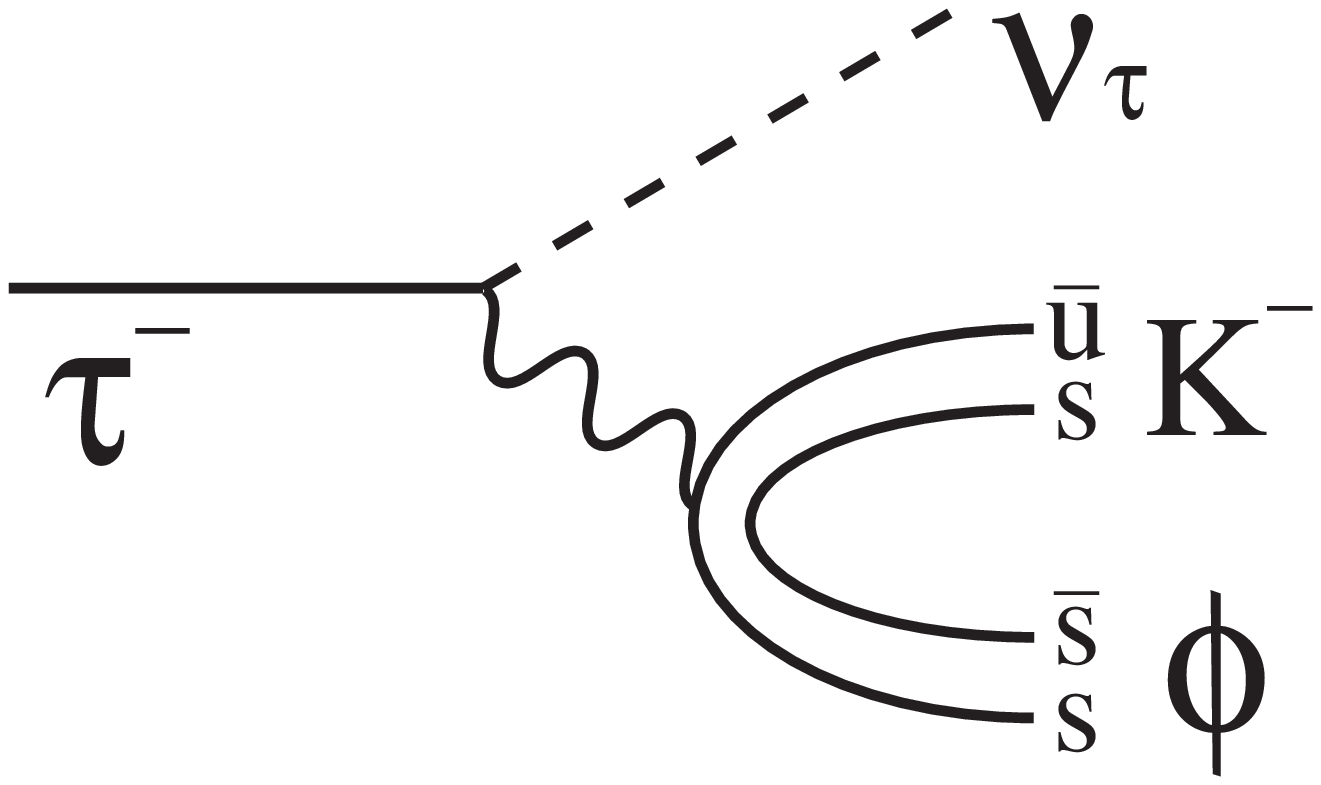}}~~~~~~
            \resizebox{3.5cm}{!}{\includegraphics{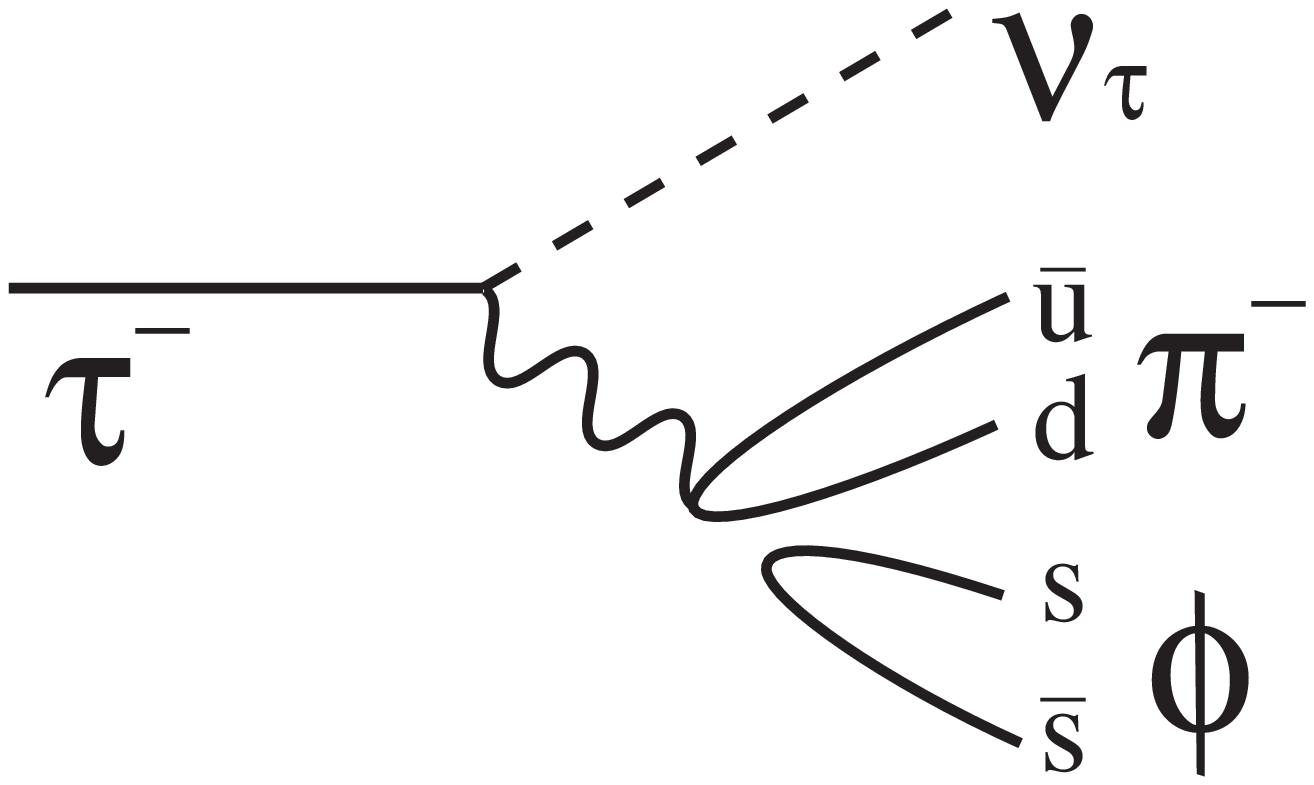}}}
\caption{Diagrams for $\tau^- \to \phi K^- \nu_\tau$ (left) and
$\tau^- \to \phi \pi^- \nu_\tau$ (right).}
\label{fig:diagram}
\end{figure}

Hadronic $\tau$ decays
with a $\phi$ meson in the final state
are valuable to investigate QCD at a low mass scale. However,
they have never been observed due to their small branching fractions.
The decay $\tau^- \to \phi K^- \nu_\tau$ is Cabibbo-suppressed and 
further restricted by its small phase space, while the decay 
$\tau^- \to \phi \pi^- \nu_\tau$ is suppressed 
by the OZI rule although it is Cabibbo-allowed
(Fig.~\ref{fig:diagram}).
The branching fraction of the former can roughly be estimated by scaling
the analogous Cabibbo-allowed decay 
$\tau^- \to K^* K^- \nu_{\tau}$~\cite{PDG} by $\tan^2 \theta_c$ and the ratio of
the phase space of the two decays, resulting in 
${\cal B}(\tau^- \to \phi K^- \nu_{\tau}) \sim 2 \times 10^{-5}$.
Similarly, the vector dominance model predicts 
${\cal B}(\tau^- \to \phi \pi^- \nu_\tau) = (1.20 \pm 0.48)\times10^{-5}$~\cite{Castro},
whereas the CVC upper limit following from the cross section
for $e^+e^- \to \phi \pi^0$ is 
${\cal B}(\tau^- \to \phi \pi^- \nu_\tau) < 3 \times 10^{-4}$ at the 90\% 
confidence level~\cite{CVC}.

Previously, the CLEO collaboration
searched for these decays using 3.1 fb$^{-1}$ of data taken on 
the $\Upsilon(4S)$ resonance. They set upper limits of
${\cal B}(\tau^- \to \phi K^- \nu_\tau)<(5.4-6.7)\times 10^{-5}$ and  
${\cal B}(\tau^- \to \phi \pi^- \nu_\tau)<(1.2-2.0)\times 10^{-4}$ at 
the 90\% confidence level, depending on the mechanism
assumed for the decay~\cite{CLEO}.
Here we report the first measurement of the 
$\tau^- \to \phi K^- \nu_\tau$ decay.
(Throughout this paper charge-conjugate states are implied.)
We also observe for the first time the decay $\tau^-\to\phi\pi^-\nu_\tau$,
but it is treated here as a background process, 
together with the kinematically 
allowed but phase-space suppressed decays
$\tau^-\to\phi \pi^- (n\pi) \nu_\tau$ ($1 \leq n\leq 4$).
The result is based on a data sample of 401 fb$^{-1}$ 
corresponding to $3.6 \times 10^8$ $\tau^+ \tau^-$ pairs
collected near the $\Upsilon(4S)$ resonance
with the Belle detector at the KEKB asymmetric-energy $e^+e^-$ 
(3.5 on 8~GeV) collider~\cite{KEKB}. 

The Belle detector is a large-solid-angle magnetic
spectrometer that
consists of a silicon vertex detector,
a 50-layer central drift chamber, an array of
aerogel threshold \v{C}erenkov counters, 
a barrel-like arrangement of time-of-flight
scintillation counters, and an electromagnetic calorimeter
comprised of CsI(Tl) crystals located inside 
a superconducting solenoid coil that provides a 1.5~T
magnetic field.  An iron flux-return located outside of
the coil is instrumented to detect $K_L^0$ mesons and identify
muons.  The detector is described in detail elsewhere~\cite{Belle}.
Two inner detector configurations were used. A 2.0 cm radius beampipe
and a 3-layer silicon vertex detector were used for the first sample
of 158 fb$^{-1}$, while a 1.5 cm radius beampipe, a 4-layer
silicon detector and a small-cell inner drift chamber were used to record  
the remaining 244 fb$^{-1}$~\cite{Ushiroda}.

\section{Event Selection}

We look for $\tau^-\to\phi K^-\nu_\tau$ candidates in 
the reaction $e^+e^-\to \tau^+\tau^-$ 
with the following signature:

\hspace*{10 mm}$\tau^-_{\rm signal} \to \phi + K^- + (\rm{missing})$ \\
\hspace*{26 mm}$\hookrightarrow K^+ K^-$ \\
\hspace*{14 mm}$\tau_{\rm tag}^+ \to (\mu/ e)^+ + n (\leq 1) \gamma + 
(\rm{missing})$, \\

\vspace{-3mm}
\noindent
where `missing' denotes other possible daughters 
not reconstructed.
The detection of $\phi$ mesons relies on the $\phi \to K^+ K^-$ decay 
(${\cal B}=(49.2\pm 0.6)\%$~\cite{PDG}); 
the final evaluation of the signal yield
is carried out using the $K^+ K^-$ invariant mass distribution. 

The selection criteria described below are determined 
from studies of Monte-Carlo (MC) simulated events.
The background samples consist of
$\tau^+\tau^-$ (1570 fb$^{-1}$, 
which does not include any decay mode
with a $\phi$ meson) and 
$q\overline{q}$ continuum, $B^0\overline{B}{}^0$, $B^+ B^-$
and two-photon processes. 
For signal, we generate samples with $2\times 10^6$
$\tau^- \to \phi K^-\nu_\tau$, 
$\phi K^-\pi^0\nu_\tau$, $\phi\pi^-\nu_\tau$ and $\phi\pi^-\pi^0\nu_\tau$
events.

The transverse momentum
for a charged track is required to be larger than 
0.06 GeV/$c$ in the barrel region ($-0.6235<\cos \theta<0.8332$, where 
$\theta$ is the polar angle 
relative to the direction opposite to that of
the incident $e^+$ beam in the laboratory frame)
and 0.1 GeV/$c$ in the endcap region ($-0.8660<\cos \theta<-0.6235$, and 
$0.8332<\cos \theta<0.9563$).
The energy of photon candidates is
required to be larger than 0.1 GeV in both regions. 

To select a $\tau$-pair sample,
we require four charged tracks in an event with zero net charge,
and a total energy of charged tracks and photons
in the center-of-mass (CM) frame less than 11 GeV.
We also require that the missing momentum in the laboratory
frame be greater than 0.1 GeV/$c$, and that its direction be within the
detector acceptance,
where the missing momentum is defined as the 
difference between the momentum of the initial $e^+e^-$
system, and the sum of the observed momentum vectors.
The event is subdivided into 3-prong and 1-prong
hemispheres according to the thrust axis 
in the CM frame. These are referred to
as the signal and tag side, respectively.
We allow at most one photon on the tag side
to account for initial state radiation,
while requiring no extra photons on the signal side
to reduce the $q\bar{q}$ backgrounds. 

We require $\cos \theta^{\rm CM}_{\rm thrust - miss}<-0.6$
to reduce backgrounds from other $\tau$ decays and
$q\bar{q}$ processes,
where $\theta^{\rm CM}_{\rm thrust - miss}$ is the opening 
angle between the thrust axis (on the signal side)
and the missing momentum in the CM frame.
In order to remove the $q\bar{q}$ background,
we require that the invariant mass of the particles on
the tag side (if a $\gamma$ is present)
be less than 1.8 GeV/$c^2$ ($\simeq m_{\tau}$).
Similarly, the effective mass of the signal side must be less
than 1.8 GeV/$c^2$.
Moreover, we require that the lepton likelihood ratio
$P_{\mu/ e}$ be greater than 0.1
for the charged track on the tag side.
Here $P_x$ is the likelihood ratio for a charged particle
of type $x$ ($x = \mu$, $e$, $K$ or $\pi$), defined as 
$P_x = L_x/(\sum_x L_x)$, 
where $L_x$ is the likelihood for particle type hypothesis $x$,
determined from responses of the relevant detectors~\cite{LID}.
The efficiencies for muon and electron identification are
92\% for momenta larger than 1.0 GeV/$c$ and
94\% for momenta larger than 0.5 GeV/$c$, respectively.

We require that both kaon daughters of the $\phi$
candidate have kaon likelihood ratios $P_K > 0.8$
and $\cos \theta >-0.6$.
The kaon identification efficiency is 82\%.
To suppress combinatorial backgrounds from other $\tau$ decays
and $q\bar{q}$ processes,
we require that the $\phi$ momentum be greater than 1.5 GeV/$c$ in the CM frame.
After these requirements, the remaining contributions
from $B^0\overline{B}{}^0$, $B^+ B^-$, Bhabha, $\mu$ pair and 
two-photon backgrounds are negligible.

To separate $\phi K^-\nu_\tau$ from $\phi\pi^-\nu_\tau$, the remaining charged 
track is required to satisfy the same kaon identification
criteria as the $\phi$ daughters.
The $\tau^+\tau^-$ and $q\bar{q}$ backgrounds
are reduced by requiring that
the opening angle ($\theta^{\rm CM}_{\phi K}$) between the
$\phi$ and $K^-$ in the CM frame 
satisfy
$\cos\theta^{\rm CM}_{\phi K}>0.92$, and that
the CM momentum of the $\phi K^-$ system
be greater than 3.5 GeV/$c$.
For $\phi \pi^-\nu_\tau$, we require that the charged track
be identified as a pion, $P_{\pi} > 0.8$,
and that the opening angle between the
$\phi$ and $\pi^-$ in the CM frame satisfy
$\cos\theta^{\rm CM}_{\phi \pi}< 0.98$.
This last requirement suppresses the background from
$\tau^- \to \phi K^-\nu_\tau$ and $\tau^- \to \phi K^-\pi^0\nu_\tau$.

\begin{figure}[htb]
\centerline{\resizebox{7cm}{!}{\includegraphics{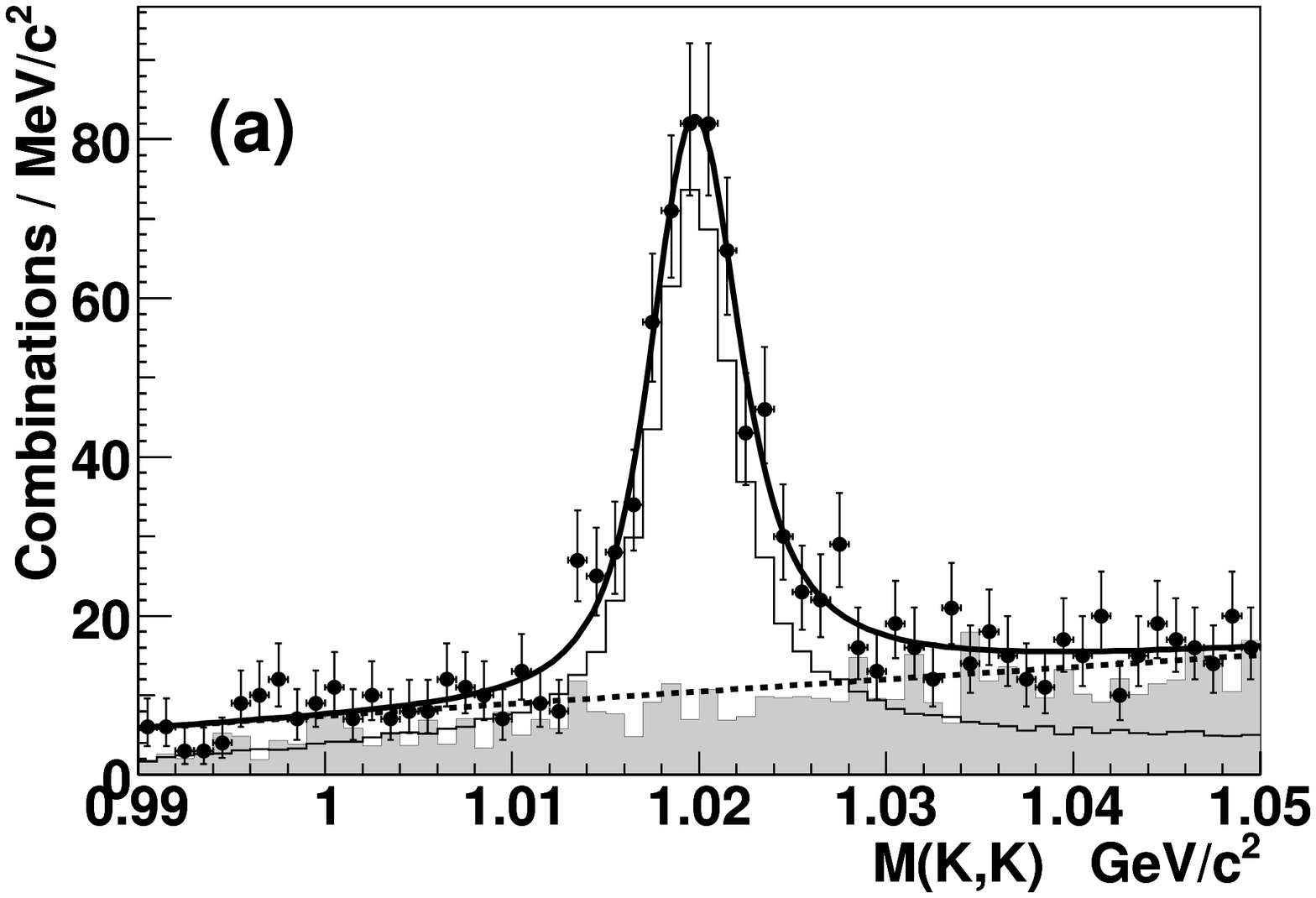}}}
\centerline{\resizebox{7cm}{!}{\includegraphics{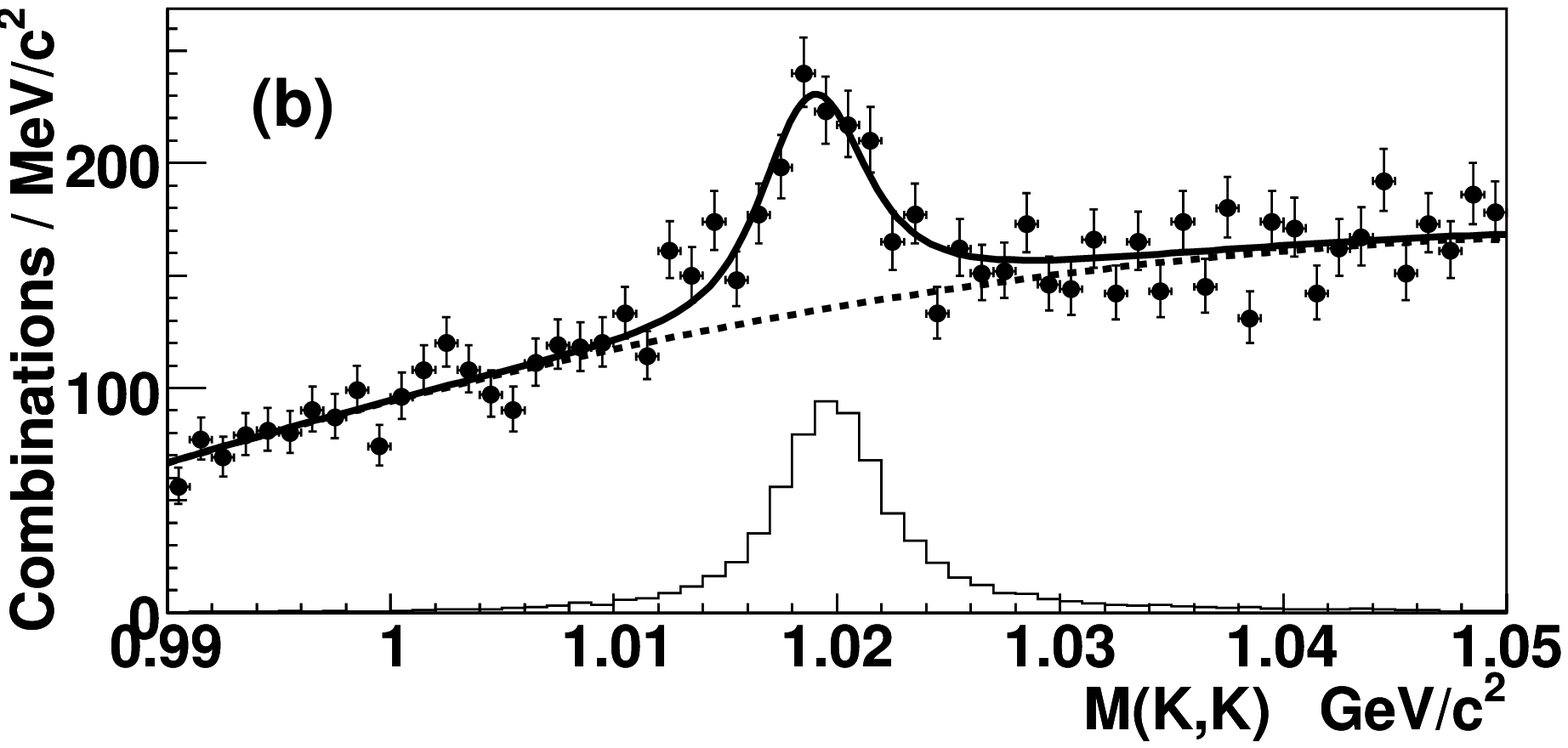}}}
\caption{$K^+ K^-$ invariant mass distributions for
(a) $\tau^- \to \phi K^- \nu_\tau$ and (b) $\tau^- \to \phi \pi^- \nu_\tau$.
Points with error bars indicate the data.
The shaded histograms show the 
expectations from $\tau^+\tau^-$ and $q\bar{q}$ background 
MC simulations.
The open histogram is the signal MC with 
${\cal B}(\tau^- \to \phi K^- \nu_\tau)=4 \times 10^{-5}$ in (a) and
${\cal B}(\tau^- \to \phi \pi^- \nu_\tau)=6 \times 10^{-5}$ in (b).
The curves show the best fit results, and
the dashed curves indicate the non-resonant background contributions.
See the text for details.}
\label{fig:fdata}
\end{figure}

Figure~\ref{fig:fdata}(a)
shows the $K^+ K^-$ invariant mass
distribution after all $\tau^- \to \phi K^-\nu_\tau$
selection requirements.
As there are two possible $K^+ K^-$ 
combinations from the $K^- K^+ K^-$ 
tracks on the signal side, this distribution has two entries per event.
Therefore, the signal MC shape includes a long tail due
to the wrong $K^+K^-$ combination.
Non-resonant backgrounds arise mainly from
$\tau^- \to K^+ K^- \pi^- \nu_\tau$, which has a branching fraction of
${\cal B} = (1.53\pm 0.10)\times 10^{-3}$~\cite{PDG}.
Small contributions are expected
from $q\bar{q}$ processes as described below.

\section{Signal and background evaluation}

The detection efficiencies $\epsilon$ for $\tau^- \to \phi K^-\nu_\tau$
and the cross-feed rates from
$\phi K^-\pi^0\nu_\tau$ and $\phi\pi^-\nu_\tau$ are evaluated, 
as listed in Table~\ref{tbl:eff}, 
from MC simulation using KKMC~\cite{KKMC},
where the $V-A$ interaction is 
assumed at the vertices and the final hadrons decay according
to non-resonant phase space.
The efficiencies include the branching fraction
for $\phi \to K^+K^-$.

\begin{table}[h]
\begin{center}
\caption{Detection efficiencies $\epsilon$ and cross-feed rates 
(\%), from MC simulation.
The errors are from the MC statistics.} 
\label{tbl:eff}
\begin{tabular}{c|ccc}\hline
  & \multicolumn{3}{c}{Decay modes} \\ 
~~Candidates~~ & $\phi K\nu$ & $\phi\pi\nu$ & $\phi K\pi^0\nu$ \\ \hline
$\tau \to \phi K\nu$ & ~~1.826$\pm$0.009~~ & ~~0.049$\pm$ 0.002~~ & 
~~0.328$\pm$0.006~~ \\
$\tau \to \phi \pi\nu$ & 0.110$\pm$0.002 & 1.663$\pm$ 0.014 & 
0.009$\pm$0.001 \\ \hline
\end{tabular}
\end{center}
\end{table}

The signal yields are extracted by a fit to the $K^+K^-$ 
invariant mass distribution. For signal, we use
a $p$-wave Breit-Wigner (BW) distribution
convolved with a Gaussian function 
(of width $\sigma$) to account for the detector resolution.
The $\phi$ width is fixed to be 
$\Gamma_{\phi} = 4.26$ MeV/$c^2$~\cite{PDG}
but $\sigma$ is allowed to float.
First- and second-order polynomial background functions
are used for $\tau^- \to \phi K^-\nu_\tau$ and $\phi\pi^-\nu_\tau$
decays, respectively.
The fit results are also shown in Fig.~\ref{fig:fdata}.
The obtained signal yields are 
$N_{\phi K\nu} = 573 \pm 32$ and 
$N_{\phi \pi\nu} = 753 \pm 84$. 
The $\sigma$'s from the fits are $1.2\pm0.3$ MeV/$c^2$
and $1.2\pm0.7$ MeV/$c^2$
for $\phi K^-\nu_\tau$ and $\phi\pi^-\nu_\tau$, respectively,
which are consistent with MC simulation.

MC studies show that only the
$\tau^- \to \phi\pi^-\nu_\tau$,
$\tau^- \to \phi K^-\pi^0 \nu_\tau$ and $q\bar{q}$ samples yield 
significant contributions
peaking at the $\phi$ mass. The contributions of other
backgrounds are less than 0.01\% and can be neglected.
The contribution of $\tau^- \to \phi\pi^-\nu_\tau$ events
to the $\phi K^- \nu_\tau$ sample is estimated using 
$N_{\phi\pi\nu}$
and the misidentification rate, as discussed below.
Other contributions are estimated as follows.

\begin{figure}[h]
\centerline{\resizebox{7cm}{!}{\includegraphics{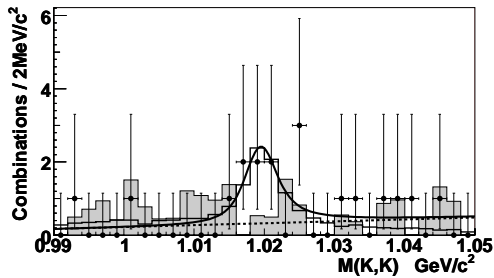}}}
\caption{$K^+ K^-$ invariant mass
distributions for $\tau^- \to \phi K^- \pi^0\nu_\tau$.
Points with error bars indicate the data.
Histograms show the MC
expectations of $\tau$-pairs (shaded) and signal (open) with
a branching fraction of $3\times 10^{-6}$.
The solid curve shows the best fit result and
the dashed curve shows the non-resonant background contribution.
}
\label{phiKpi0}
\end{figure}

To evaluate the branching fraction and background
contribution from $\tau^- \to \phi K^-\pi^0\nu_\tau$,
we select $\pi^0 \to \gamma\gamma$ candidates and combine them with
$\phi K^- \nu_\tau$ combinations that satisfy the requirements listed above.
The signal yield is estimated by fitting the resulting
$K^+ K^-$ invariant mass distribution with 
a $p$-wave BW distribution plus a linear background function,
as shown in Fig.~\ref{phiKpi0}.
The resulting yield is $8.2\pm3.8$ $\phi K \pi^0 \nu$ events.
Using a detection efficiency $\epsilon_{\phi K\pi^0\nu} = 
(0.396\pm 0.007)\%$ obtained from MC simulation,
and an $e^+e^- \to \tau^+\tau^-$
sample normalization 
$N_{\tau\tau} = 401\,\mathrm{fb}^{-1} \times 0.892\,\mathrm{nb}
= 3.58 \times 10^{8}$, we obtain a branching fraction
${\cal B}(\tau^- \to \phi K^-\pi^0\nu_\tau) = (2.9\pm1.3)\times 10^{-6}$.
However, this must be corrected for the unknown contamination of
$\tau^- \to \phi \pi^- \pi^0 (n\pi) \nu_\tau$ $(0\leq n \leq 3)$
decays.
Using this value, 
we estimate the $\tau^- \to \phi K^-\pi^0\nu_\tau$ 
background in
the $\tau^- \to \phi K^-\nu_\tau$ sample to be
$N_{\phi K\pi^0\nu} = (6.8\pm 3.1)$ events,
given a cross-feed rate 
for $\tau^- \to \phi K^-\pi^0\nu_\tau$ to the
$\tau^- \to \phi K^-\nu_\tau$ sample of
$(0.328\pm 0.006)\%$ (see Table~\ref{tbl:eff}). 

From a MC study, we find a $q\bar{q}$
contamination of $N_{q\bar{q}} = 6.6\pm 2.5$.
To take into account the uncertainty in $\phi$ production
in the $q\bar{q}$ MC,
we compare MC results with enriched $q\bar{q}$ data by demanding 
that the effective 
mass of the tag side be larger than 1.8 GeV/$c^2$. 
With this selection, the background is $q\bar{q}$ dominated and
the other backgrounds are negligible. 
The yield in data is $262\pm 21$ events, and
the yield in the $q\bar{q}$ MC is $117\pm 10$ events.
We subsequently scale the above $q\bar{q}$
background estimate by the ratio $f = 2.23 \pm 0.26$;
the result is $N_{q\bar{q}} = 14.8\pm 5.8$ events.

\section{Results}

The peaking backgrounds described above, 
$\tau^- \to \phi K^-\pi^0\nu_\tau$ and $q\bar{q}$, are subtracted 
from the signal yield, leaving
$N_{\phi K\nu} = (573\pm32)-(6.8\pm3.1)-(14.8\pm
5.8) = 551\pm33 ~~{\rm events}$.
To take into account cross-feed between 
$\tau^- \to \phi K^-\nu_\tau$ and $\tau^- \to \phi\pi^-\nu_\tau$
due to particle misidentification ($K\leftrightarrow \pi$), 
we solve the following simultaneous equations:
\begin{eqnarray}
N_{\phi K\nu} &=& 2 N_{\tau\tau} \left( 
\epsilon_{\phi K\nu} \times {\cal B}_{\phi K\nu} + 
\epsilon_{\phi \pi\nu}^{\phi K\nu} \times {\cal B}_{\phi \pi\nu} 
\right), \\
N_{\phi \pi\nu} &=& 2 N_{\tau\tau} \left( 
\epsilon_{\phi K\nu}^{\phi\pi\nu} \times {\cal B}_{\phi K\nu} + 
\epsilon_{\phi \pi\nu} \times {\cal B}_{\phi \pi\nu}
\right), 
\end{eqnarray}
where ${\cal B}_{\phi K\nu}$ and ${\cal B}_{\phi \pi\nu}$ are
the branching fractions for 
$\tau^- \to \phi K^- \nu_\tau$ and $\tau^- \to \phi \pi^- \nu_\tau$, 
respectively. 
The detection efficiencies, $\epsilon$'s,
are listed in Table~\ref{tbl:eff}.
The factor $\epsilon_{\phi K\nu}^{\phi\pi\nu}$
is the efficiency for reconstructing
$\tau^- \to \phi K^-\nu_\tau$ as 
$\tau^- \to \phi\pi^-\nu_\tau$ while
$\epsilon_{\phi \pi\nu}^{\phi K\nu}$
is the efficiency for reconstructing
$\tau^- \to \phi \pi^-\nu_\tau$ as $\tau^- \to \phi K^-\nu_\tau$. 
The resulting branching fraction for $\tau^- \to \phi K^-\nu_\tau$ is 
\begin{eqnarray}
{\cal B}_{\phi K\nu} = (4.05 \pm 0.25) \times 10^{-5},
\end{eqnarray}
where the uncertainty is due to the statistical 
uncertainty in the $N_{\phi K\nu}$ and $N_{\phi\pi\nu}$ terms.
The uncertainty in the detection efficiencies, 
$\epsilon$'s, will be taken into account in the systematic error.
The result for ${\cal B}_{\phi \pi\nu}$ is
${\cal B}_{\phi \pi\nu} = (6.05 \pm 0.71) \times 10^{-5}$;
however, small background from
$\tau^- \to \phi \pi^- (n\pi) \nu_\tau$ $(1\leq n \leq 4)$
decays is included and must be subtracted to obtain 
the final branching fraction.

The systematic uncertainties are estimated as follows: 
The uncertainties in
the integrated luminosity, 
$\tau^+\tau^-$ cross-section and trigger efficiency are 
1.4\%, 1.3\% and 1.1\%, respectively. 
Track finding efficiency has an uncertainty of 4.0\%. 
Uncertainties in lepton and kaon identification efficiencies and fake rate
are evaluated, respectively, to be 3.2\% and 3.1\% by averaging the
estimated uncertainties depending on momentum and polar angle of 
each charged track. 
To evaluate the systematic uncertainty of fixing $\Gamma_{\phi}$ in the BW fit, 
we calculate the change in the signal yield when
$\Gamma_{\phi}$ is varied
by $\pm 0.05$ MeV/$c^2$
(the uncertainty quoted by the PDG)~\cite{PDG}:
the result is $0.2\%$. 
The branching fraction for $\phi \to K^+ K^-$ gives an 
uncertainty of $1.2\%$~\cite{PDG}.
The signal detection efficiency 
$\epsilon_{\phi K\nu}$ has an uncertainty of 0.5\%
due to MC statistics. 
A total systematic uncertainty of 6.5\% is obtained 
by adding all uncertainties in quadrature. 
The resulting branching fraction is then 
\begin{eqnarray}
{\cal B}(\tau^- \to \phi K^-\nu_\tau) = (4.05 \pm 0.25 \pm 0.26) \times 10^{-5}.
\end{eqnarray}

\begin{figure}[htb]
\centerline{\resizebox{8cm}{!}{\includegraphics{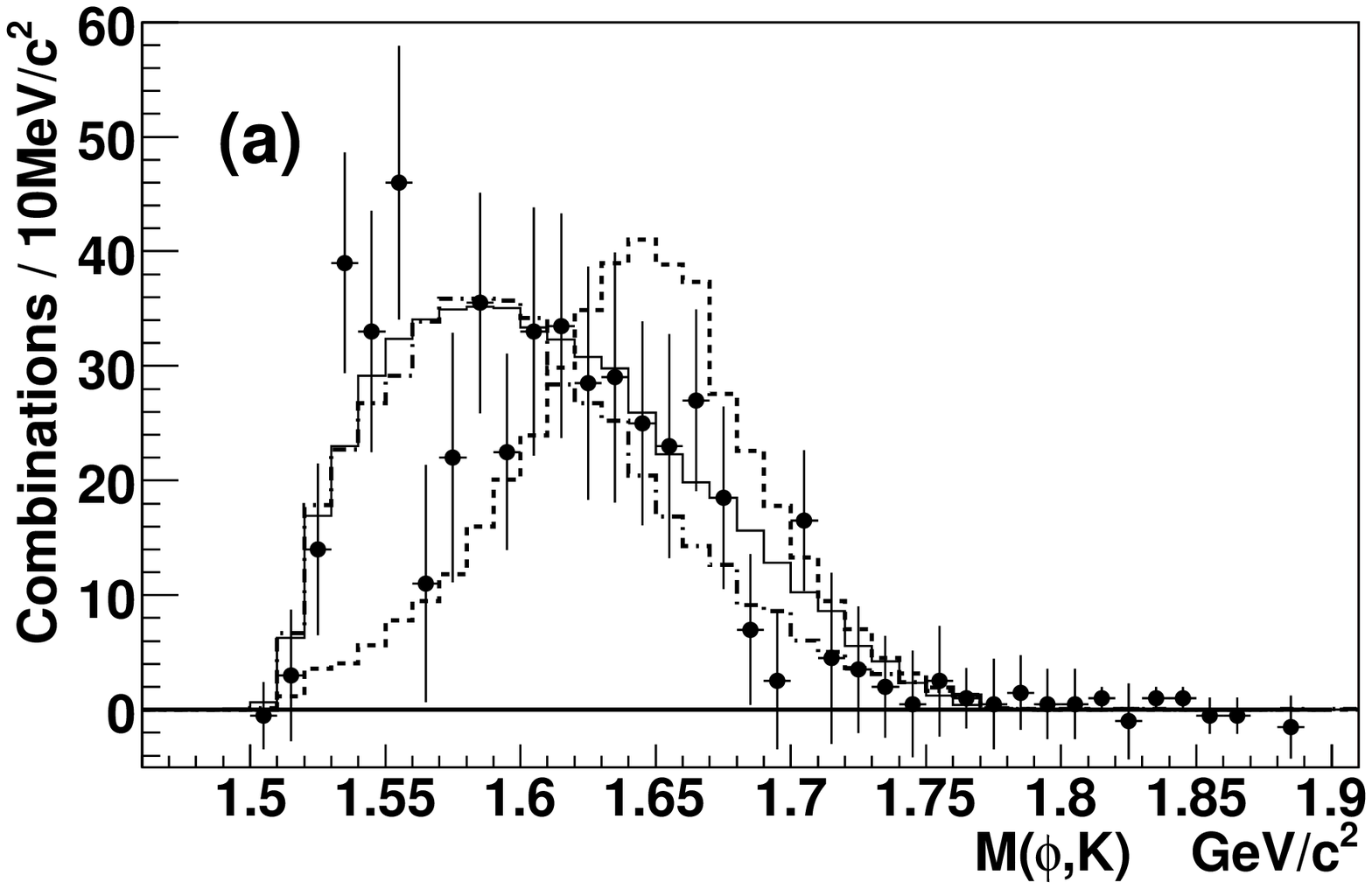}}}
\centerline{\resizebox{8cm}{!}{\includegraphics{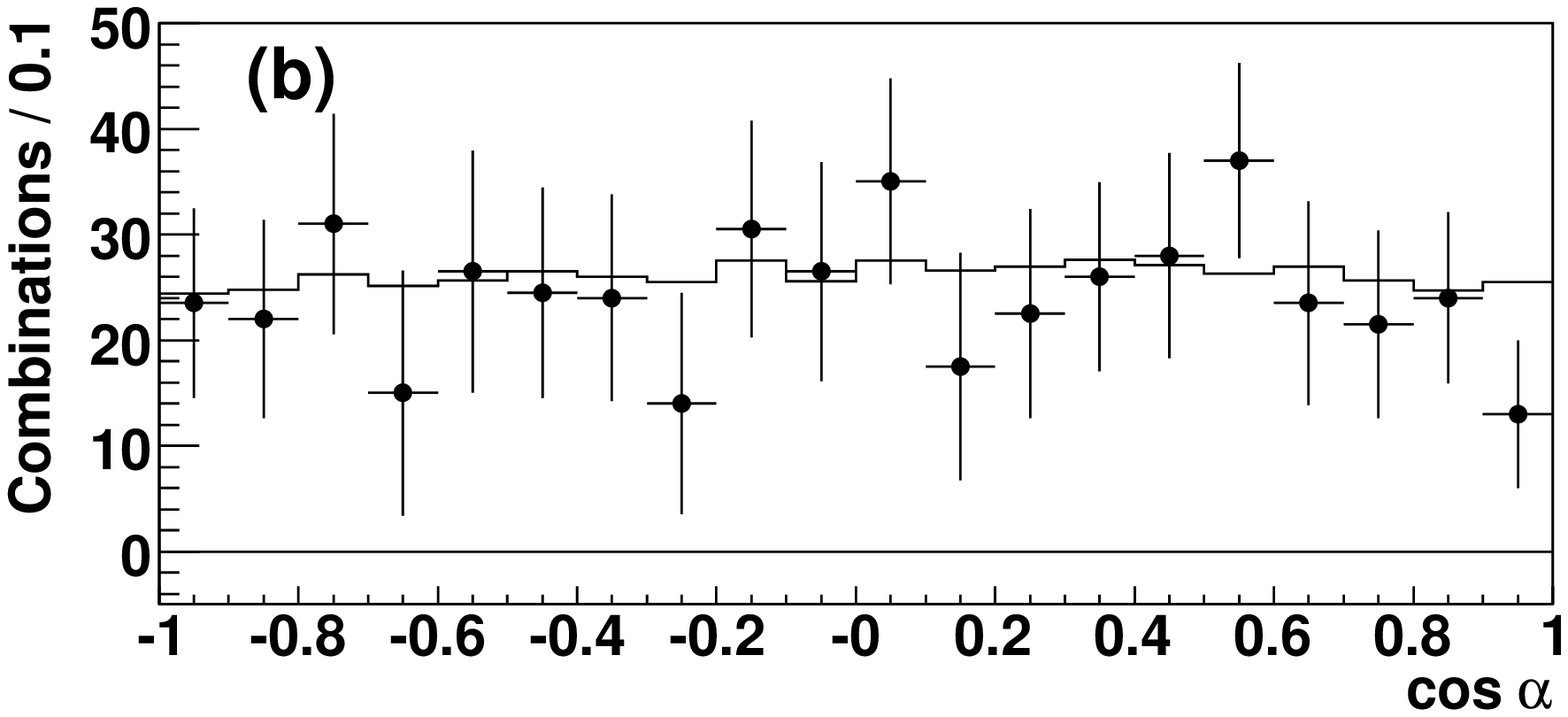}}}
\caption{(a) invariant mass and (b) angular distributions for 
the $\phi K^-$ system. The non-$\phi$-resonant
backgrounds are subtracted using the sideband spectra.
Points with error bars indicate the data. 
The open histogram shows the phase space distributed signal MC, and 
dotted and dot-dashed histograms indicate the signal MC mediated by a resonance 
with $M=1650$ MeV/$c^2$ and $\Gamma=100$ MeV/$c^2$ and 
$M=1570$ MeV/$c^2$ and $\Gamma=150$ MeV/$c^2$, respectively.
In the MC, a branching fraction of $4\times 10^{-5}$ is assumed.
(b) $\phi$'s angular distribution in the $\phi K^-$ rest frame,
where the negative of the lab frame direction in the
$\phi K^-$ frame is taken as the reference axis.
}
\label{fig:pm3}
\end{figure}

Finally, we consider the possibility that a resonant
state contributes to the final $\phi K^-$ hadronic system. 
We generate a resonant MC with the KKMC simulation program. 
The weak current is generated with a $V-A$ form while
the $\phi K^-$ system is assumed to be produced from a 2-body decay of a
resonance.
In Fig.~\ref{fig:pm3}(a), the $\phi K^-$ mass distribution 
for data is compared to MC;
the combinatorial background is subtracted using the $K^+ K^-$ sideband.
The MC distributions correspond to 
($M, \Gamma$) =(1650, 100) MeV/$c^2$, ($M, \Gamma$) =(1570, 150) MeV/$c^2$,
and also non-resonant phase space.
Figure~\ref{fig:pm3}(b) shows the $\phi$'s angular distribution
in the $\phi K^-$ rest frame ($\cos \alpha$),
where the negative of the lab frame direction in the
$\phi K^-$ frame is taken as the reference axis.
It indicates an isotropic distribution in the $\phi K^-$ system. 
For both the invariant mass and angular distributions of 
the $\phi K^-$ system, 
the phase space MC reproduces the signal distribution well.
We therefore neglect systematic uncertainty due to
possible resonant structure.
On the other hand, the 1650 MeV/$c^2$ state
assumed in the CLEO search~\cite{CLEO},
indicated by the dotted histogram in Fig.~\ref{fig:pm3}(a), 
clearly cannot account for the entire signal.
If production via a single resonance is assumed,
the best agreement with data
is found for a mass and a width of
$\simeq$1570 MeV/$c^2$ and $\simeq$150 MeV/$c^2$, respectively, as shown 
by the dot-dashed histogram. 
However, since the shape of the resonant MC is similar to
the phase-space-distributed MC,
we cannot draw any strong conclusions about
an intermediate resonance with 
$\Gamma \sim O(100 {\rm MeV/}c^2)$
in this narrow mass range of $\sim$250 MeV/$c^2$.

\section{Conclusion}

Using 401 fb$^{-1}$ of data, we make the first 
observation of the rare decay $\tau^- \to \phi K^-\nu_\tau$.
The measured branching fraction is
\begin{eqnarray}
{\cal B}(\tau^- \to \phi K^-\nu_\tau) = (4.05 \pm 0.25 \pm 0.26) \times 10^{-5}.
\end{eqnarray}

\smallskip
\bigskip
\noindent
{\bf Acknowledgments}
\smallskip

We gratefully acknowledge the essential
contributions of Mari Kitayabu, which are described in her bachelor thesis 
at Nagoya University.
We thank the KEKB group for the excellent operation of the
accelerator, the KEK cryogenics group for the efficient
operation of the solenoid, and the KEK computer group and
the National Institute of Informatics for valuable computing
and Super-SINET network support. We acknowledge support from
the Ministry of Education, Culture, Sports, Science, and
Technology of Japan and the Japan Society for the Promotion
of Science; the Australian Research Council and the
Australian Department of Education, Science and Training;
the National Science Foundation of China and the Knowledge
Innovation Program of the Chinese Academy of Sciences under
contract No.~10575109 and IHEP-U-503; the Department of
Science and Technology of India; 
the BK21 program of the Ministry of Education of Korea, 
the CHEP SRC program and Basic Research program 
(grant No.~R01-2005-000-10089-0) of the Korea Science and
Engineering Foundation, and the Pure Basic Research Group 
program of the Korea Research Foundation; 
the Polish State Committee for Scientific Research; 
the Ministry of Science and Technology of the Russian
Federation; the Slovenian Research Agency;  the Swiss
National Science Foundation; the National Science Council
and the Ministry of Education of Taiwan; and the U.S.\
Department of Energy.

\end{document}